\documentclass[printer]{aa}

\bibpunct{(}{)}{;}{a}{}{,}

\usepackage{lscape}
\usepackage{graphicx}
\usepackage[varg]{txfonts}
\newcommand{\hii}{H\textsc{ii}}
\def\ks{km s$^{-1}$}
\def\d{$^\circ$}

\def\s{$^{\prime\prime}$}

\def\cm3{cm$^{-3}$}

\def\2{$^{12}$CO}
\def\3{$^{13}$CO}
\def\8{C$^{18}$O}

\def\cm2{cm$^{-2}$}
\begin{document}

\title{Exploring the $^{13}$CO/C$^{18}$O abundance ratio towards Galactic young stellar objects and H\textsc{ii} regions}
%in $27\fdg5\leq l\leq46\fdg5$ and $\left|b\right|\leq 0\fdg5$}
\author {M. B. Areal \inst{1}
\and S. Paron \inst{1,2}
\and M. Celis Pe\~{n}a \inst{1}
\and M. E. Ortega \inst{1}
}

\institute{CONICET-Universidad de Buenos Aires. Instituto de Astronom\'{\i}a y F\'{\i}sica del Espacio
             CC 67, Suc. 28, 1428 Buenos Aires, Argentina 
\and Universidad de Buenos Aires. Facultad de Arquitectura, Dise\~{n}o y Urbanismo. Buenos Aires, Argentina
}

\offprints{M. B. Areal}

   \date{Received <date>; Accepted <date>}

\abstract{}{Determining molecular abundance ratios is important not only for the study of the Galactic chemistry but also
because they are useful to estimate physical parameters in a large variety of interstellar medium environments.
The CO is one of the most important molecules to trace the molecular gas in the interstellar medium, and 
the \3/\8 abundance ratio is usually used to estimate molecular masses and densities of regions with moderate to high
densities. Nowadays this kind of isotopes ratios are in general indirectly derived from elemental abundances ratios. 
We present the first \3/\8 abundance ratio study performed from CO isotopes 
observations towards a large sample of Galactic sources of different nature at different locations.}
{To study the \3/\8 abundance ratio it was used \2 J=3--2 data obtained form the CO High-Resolution Survey, 
\3 and \8 J=3--2 data from the \3/\8 (J=3--2) Heterodyne Inner Milky Way Plane Survey, and 
some complementary data extracted from the James Clerk Maxwell Telescope database. It was analyzed a sample of 198 sources composed by young stellar 
objects (YSOs), \hii~and diffuse \hii~regions as catalogued in the Red MSX Source Survey in $27\fdg5\leq l\leq46\fdg5$ and $\left|b\right|\leq 0\fdg5$.}
{Most of the analyzed sources are located in the galactocentric distance range 4.0--6.5 kpc. 
We found that YSOs have, in average, smaller \3/\8 abundance ratios than \hii~and diffuse \hii~regions. Taking into 
account that the gas associated with YSOs should be
less affected by the radiation than in the case of the others sources, selective far-UV
photodissociation of \8 is confirmed.
The \3/\8 abundance ratios obtained in this work are systematically lower than the predicted from 
the known elemental abundance relations.
These results would be useful in future studies of molecular gas related to YSOs and \hii~regions based on the observation of these isotopes.
}
{}

\titlerunning{$^{13}$CO/C$^{18}$O abundance ratio towards YSOs and H\textsc{ii} regions}
\authorrunning{M.B. Areal et al.}

\keywords{ISM: abundances -- ISM: molecules -- Galaxy: abundances --  {\it (ISM:)} H\textsc{ii} regions --  Stars: formation   }

\maketitle

\section{Introduction}

Studying molecular abundances towards different Galactic sources and their association with the the surrounding interstellar medium 
is a very important issue in astrophysics. To study the chemical evolution of the Galaxy, and hence the physical processes 
related to the chemistry, is crucial to know accurate values of molecular abundances. 

It is well known that the carbon monoxide is the second most abundant molecule in the universe, and the rotational transitions of 
$^{12}$C$^{16}$O (commonly \2) are easily observable. Our knowledge of the molecular gas distribution along 
the Galaxy comes mainly from the observation of this molecular species and its isotopes, such as $^{13}$C$^{16}$O (\3)
and $^{12}$C$^{18}$O (\8). It can be affirmed that any study involving molecular gas traced by the CO  
isotopes needs molecular abundance relations to derive physical and chemical parameters.

\citet{wilson94} and \citet{wilson99} presented one of the most complete and used works about interstellar abundances focused 
on chemical elements. Regarding to C and 
O, and hence to CO and its isotopes, most of the works in the literature that have to assume an abundance ratio concerning to the CO use the
values from the Wilson's papers. \citet{wilson94} have shown that the $^{12}$C/$^{13}$C and the $^{16}$O/$^{18}$O ratios depend on the distance
to the Galactic center. These relations were obtained mainly from H$_{2}$CO absorption lines observations because, as the authors mention, direct 
measurements from the CO presented some problems such as a lack of surveys with a complete set of CO isotopes with good S/N ratio. 
Later, \citet{milam05} studied the $^{12}$C/$^{13}$C ratio through the
N = 1--0 transition of the CN radical and also found the same dependency with the galactocentric distance. 
%In addition, the authors show a good 
%agreement between the $^{12}$C/$^{13}$C ratios obtained from CN, CO, and H$_{2}$CO.
Besides this dependency with the distance to the Galactic center, the isotope abundance ratios can present variations along the same molecular cloud. 
This is the case for the \2/\3 abundance ratio, which may vary considerably within the same molecular cloud due to chemical fractionation and 
isotope-selective chemical processes (see \citealt{szucs14} and references therein). 

Concerning to the \3/\8 abundance ratio, early works of \citet{dick79} and \citet{lang80} have shown spatial gradients from the edge to the
center of molecular clouds. More recently, direct observations of \2, \3, and \8 lines have been used to determine abundance ratios and its relation 
with far-UV radiation towards different regions in molecular clouds \citep{lin16,kong15,shima14}, showing that the \3/\8 abundance ratio may exhibit 
significant spatial variations. These kind of studies show that the abundance ratios can depend not only on galactocentric distances, but 
also on the type of the observed source and its surroundings. \hii~regions and young stellar objects (YSOs) can be useful targets to study this 
issue because the molecular gas at their surroundings are affected by far-UV radiation, jets, winds and outflows. Studies towards a large 
sample of these objects using the CO J=3--2 line, which has a critical density 
$\gtrsim$ 10$^{4}$ cm$^{-3}$ and nowadays is extensively used to observe their surroundings, are needed. 
%Moreover, CO abundance studies with lines others
%than the J=1--0 are useful to compare with results from theoretical models about the influence of photodissociation rates in the molecule 
%rotational levels \citep{warin96}.

It is important to mention that if one needs to assume some abundance ratio between molecular
isotopes, the best would be to use a value that have been derived directly from the molecular species instead from the ratio
between the elements that compose the molecules. In the case of the \3/\8 abundance ratio most of the works in the
literature perform a double ratio from the Wilson's $^{12}$C/$^{13}$C and $^{16}$O/$^{18}$O expressions, due to, as mention above, a lack of
CO isotopes surveys with good S/N that cover large areas in the Galaxy.

At present there are some surveys of \2, \3 and \8 J=3--2 with good S/N ratio that allow us to perform abundances estimates towards many sources 
of different nature in the Galaxy. We present here a study of the \3/\8 abundance ratio 
towards a large sample of YSOs and \hii~regions in a region of about 20\d$\times$1\d~at the first Galactic quadrant. This is the first 
large survey of \3/\8 abundance ratios
and it was performed  with the aims of testing the known CO abundances relations using a modern data set, and to explore how the abundance 
ratios depends not only on the distance to the Galactic center but also on the type of source or region observed.

\begin{figure*}[tt!]
\centering
\includegraphics[width=9.1cm]{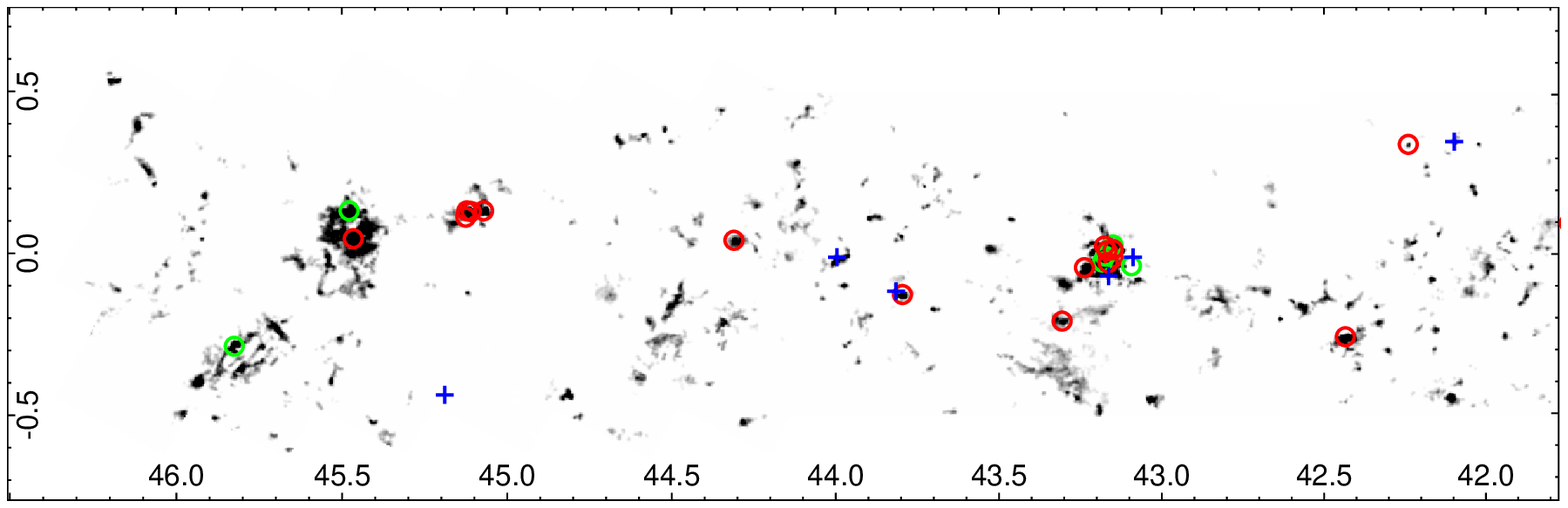}
\includegraphics[width=9.1cm]{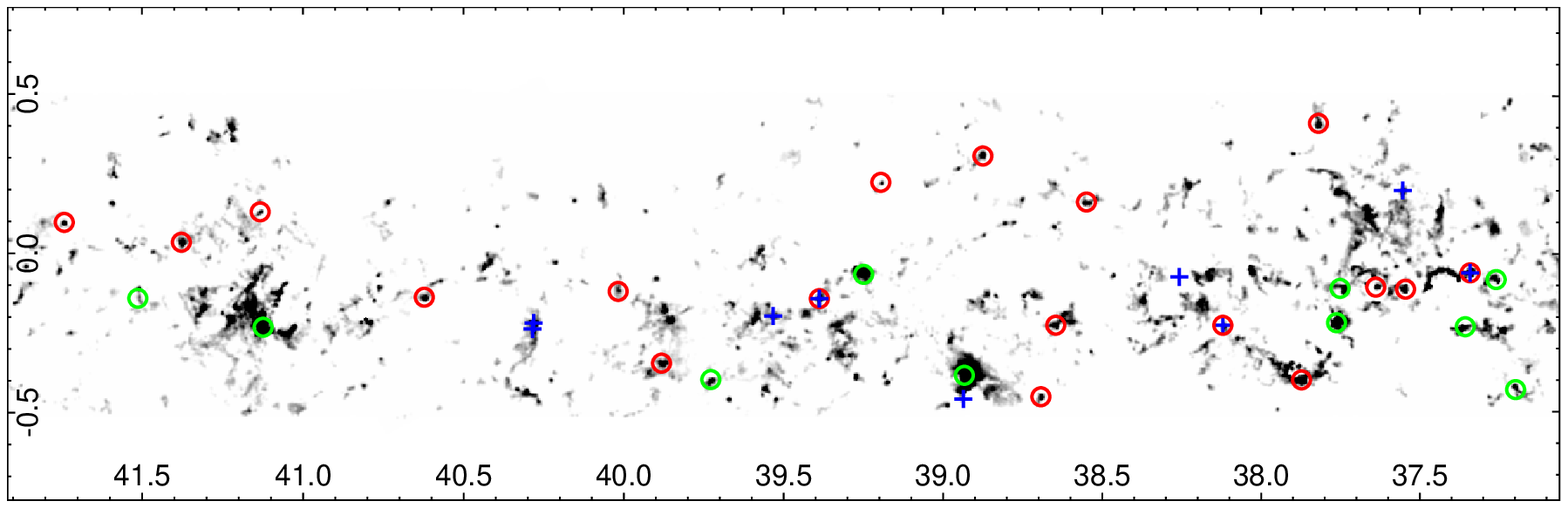}
\includegraphics[width=9.1cm]{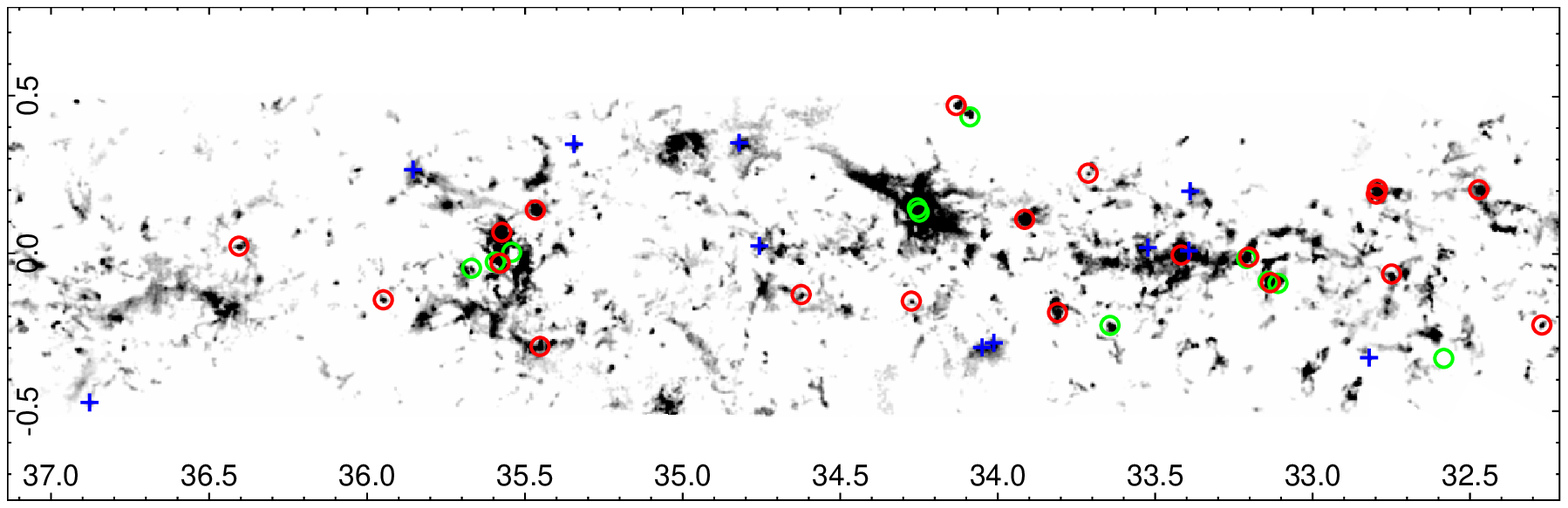}
\includegraphics[width=9.1cm]{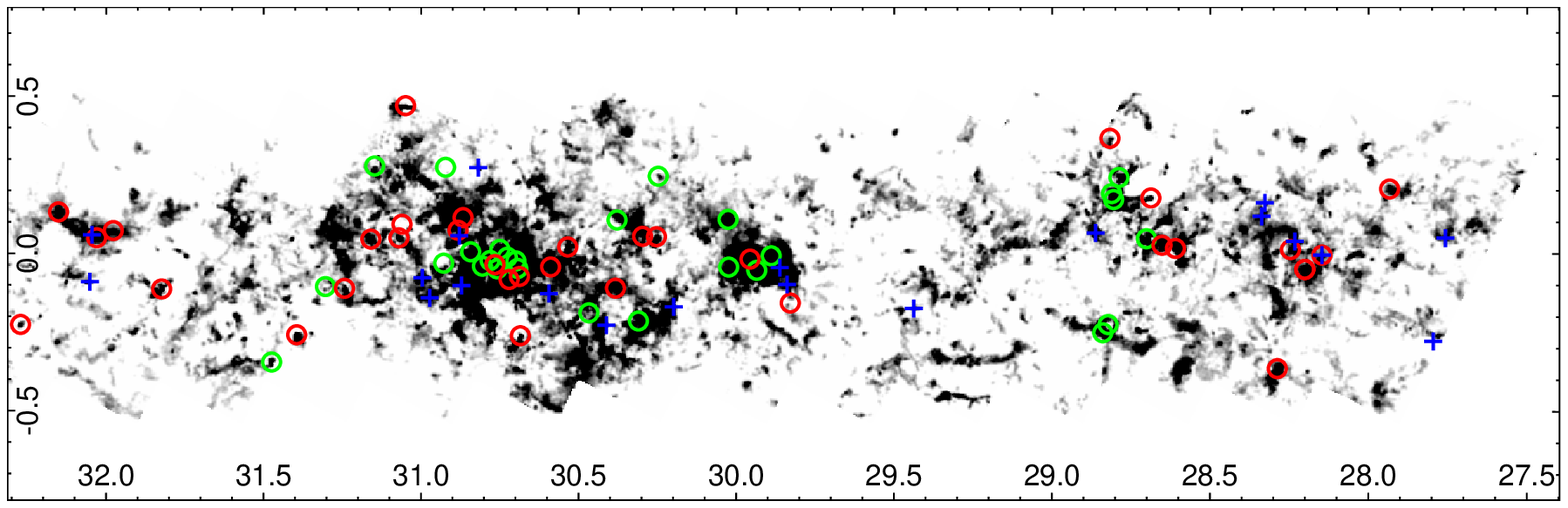}
\caption{Integrated \3 J=3--2 emission maps of the whole region surveyed by the CHIMPS survey.
All selected sources from the Red MSX Source Catalog \citep{lums13} lying in this region are presented as follows: YSOs (blue crosses),
\hii~regions (red circles), and diffuse \hii~regions (green circles).}
\label{galaxy}
\end{figure*}

\section{Data and sources selection}

The data of the CO isotopes were extracted from two public databases performed with the 15 m James Clerck Maxwell 
Telescope (JCMT) in Hawaii.
The \2 J=3--2 data were obtained from the CO High-Resolution Survey (COHRS) with an angular and spectral resolution of
14\s~and 1 \ks~(see \citealt{dempsey13}). The data of the other CO isotopes were obtained from the
\3/\8 (J=3--2) Heterodyne Inner Milky Way Plane Survey (CHIMPS), which have an angular and spectral resolution of
15\s~and 0.5 \ks~\citep{rigby16}. The intensities of both set of data are on the $T_{A}^{*}$ scale, and it 
was used the mean detector efficiency $\eta_{\rm mb} = 0.61$ for the \2, and $\eta_{\rm mb} = 0.72$ for the \3 and \8 
to convert $T_{A}^{*}$ to main beam brightness temperature (T$_{\rm mb} = T_{A}^{*}/\eta_{\rm mb}$)  \citep{buckle09}.

Taking into account that the CHIMPS survey covers the Galactic region in $27\fdg5\leq l\leq46\fdg5$ and $\left|b\right|\leq 0\fdg5$,
we selected all sources catalogued as YSOs, \hii~and diffuse \hii~regions lying in this area from the Red MSX Source Survey \citep{lums13}, 
which is the largest statistically selected catalog of young massive protostars and \hii~regions to date. Figure\,\ref{galaxy} shows
the 198 sources among YSOs (blue crosses), \hii~regions (red circles), and diffuse \hii~regions (green circles). 
This source classification, and the separation from other kind of sources, was done by \citet{lums13} using several multiwavelength 
criteria. The authors combined near- and mid-infrared color criteria, analysis of spectral energy distributions, near-infrared 
spectroscopy, analysis of radio continuum fluxes and maser emission, and comparisons with  other published lists of sources and catalogs 
to classify the sources. In addition, the morphology of the
emission at different wavelengths was also analyzed. For example sources with stronger emission at 12 $\mu$m than at either 8 $\mu$m or 14 $\mu$m
are often extended and classified as diffuse \hii~regions. \citet{lums13} remark that the final source 
classification was decided individually for every source. What it is important for our work is that the classification presented 
in the Red MSX Source Survey differentiates efficiently between two classes of stellar objects: the youngest (YSOs) and the
more evolved (\hii~regions). The `\hii~region' category can include ultracompact, compact, and point-like
\hii~regions, while the `diffuse \hii~regions' category refers to extended and likely more evolved \hii~regions.

At the Galactic longitude range covered by CHIMPS the COHRS survey is restricted to Galactic latitudes of $\left|b\right|\leq 0\fdg25$. Thus
the \2 data for sources lying in $\left|b\right|\geq 0\fdg25$ and $\left|b\right|\leq 0\fdg5$ were obtained from the JCMT 
database\footnote{http://www.cadc-ccda.hia-iha.nrc-cnrc.gc.ca/en/jcmt/}. In these cases it was used the reduced data.

The spectra of each isotope were extracted from the position of each catalogued source in the Red MSX Source Survey. 
Given that we need the emission from the three isotopes to obtain the \3 and \8 column densities (N(\3) and N(\8))
and then the abundance ratio (X$^{13/18}=$N(\3)/N(\8)), 
in the cases of sources that some isotope does not present emission above the noise level in the surveys, 
we checked in the JCMT database if there are observing programs around the source coordinate other than those used 
to perform the surveys. In the affirmative case we investigated these data in order to find the lacking spectra.

The data were visualized and analyzed with the Graphical Astronomy and Image Analysis Tool (GAIA)\footnote{GAIA is a derivative of 
the Skycat catalogue and image display tool, developed as part of the VLT project at ESO. Skycat and GAIA are free software under 
the terms of the GNU copyright.} and with tools from the Starlink software package \citep{currie14} such as the Spectral Analysis Tool
(Starlink SPLAT-VO). The typical rms noise levels of the spectra, in units of $T_{A}^{*}$, are: 0.25, 0.35, and 0.40 K for 
the \2, \3, and \8, respectively.

\section{Results}

From the sample of 198 sources lying in the analyzed region (see Fig.\,\ref{galaxy}), we obtained spectra from the 
three CO isotopes in 114 cases.
Table\,1-at the end of the manuscript- presents the line parameters obtained from Gaussian fits to the spectra of each CO isotope, including also the 
velocity-integrated line emission in the case of \3 and \8. The assigned number, 
the source designation and its classification 
from the Red MSX Source Survey are included in Cols.\,1, 2, and 3. For simplicity, errors are not included in the Table. 
In the case of the \2 and \3 the typical errors  (the formal 1$\sigma$ value for the model of the Gaussian line shape) in T$_{mb}$ 
are between 5 and 10 \%, while the typical error in this parameter for the \8 ranges from 10 to 20 \%. The integrated
line emission has typical errors of 5--10 \%, and 10--20 \%~for the \3 and \8, respectively.
All the \8, and most of the \3 spectra, present only one component along
the velocity axis, which represents the emission related to the catalogued source, defining in that way its central velocity (v$_{LSR}$). 
In several cases this velocity could be checked with methanol and/or ammonia maser emission catalogued in the Red MSX Source Survey. 
When the \2 spectrum of any source presents several velocity components, it was selected the one coinciding with the v$_{LSR}$ 
measured from the other isotopes.

Given that the galactocentric distance of the sources is an 
important parameter to take into account to study abundances \citep{wilson94}, it is also included in Table\,1-at the end of the manuscript- (Col.\,4).
Most of the sources have catalogued distances to us in the Red MSX Source Survey, which were used to estimate the corresponding 
galactocentric distance. In the cases that it is not included a distance in the Red MSX Source Survey, we derived it from the \8 central 
velocity (v$_{LSR}^{18}$) using the Galactic rotation model of \citet{fich89}, obtaining
a pair of possible distances to us (the nearest and farthest) due to the distance ambiguity in the first Galactic quadrant. Finally, using 
this pair of possible distances to us it was obtained the corresponding galactocentric distances, which in all cases, are almost the same 
independently if we use the nearest or farthest distances derived from the v$_{LSR}^{18}$. Thus, no ambiguity in the galactocentric 
distances appears in the studied sources. Most of the sources are located at galactocentric distances between 4.0 and 6.5 kpc, which taking 
into account the Galactic longitude range of the surveyed area, indicates that we are mainly studying sources in the Scutum-Crux 
and Sagitarius-Carina Galactic arms.

\begin{figure}[h]
\centering
\includegraphics[width=7cm]{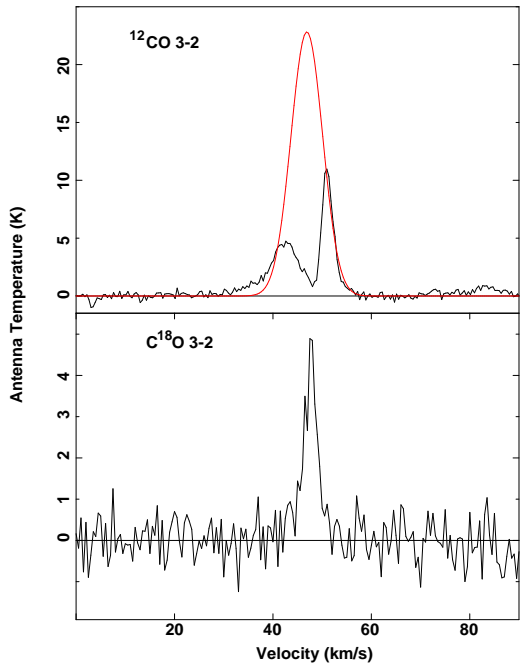}
\caption{Example of a \2 self-absorption correction.
The \8 emission peaks at the \2 dip, showing that the \2 emission is self-absorbed. It is shown the
Gaussian fit to the central component corrected for absorption.}
\label{absCorrect}
\end{figure}

We estimated the column densities of \3 and \8 on the assumption that the rotational levels of these molecules
are in local thermodynamic equilibrium (LTE). The optical depths ($\tau_{\rm ^{13}CO}$ and $\tau_{\rm C^{18}O}$) and column densities 
(N(\3) and N(\8)) can be derived using the following equations:
\begin{equation}
  \tau_{\rm ^{13}CO} = - ln\left(1 - \frac{T_{mb}({\rm ^{13}CO})}{15.87\left[\frac{1}{e^{15.87/T_{ex}}-1} - 0.0028\right]}\right)    
\label{tau13}
\end{equation}

\begin{equation}
{\rm N(^{13}CO)} = 8.28 \times 10^{13}~e^{\frac{15.85}{T_{ex}}}\frac{T_{ex} + 0.88}{1 - e^{\frac{-15.87}{T_{ex}}}} 
\int{\tau_{\rm ^{13}CO}{\rm dv}}
\label{N13}
\end{equation}
with 
\begin{equation}
\int{\tau_{\rm ^{13}CO}{\rm dv}} = \frac{1}{J(T_{ex}) - 0.044} \frac{\tau_{\rm ^{13}CO}}{1-e^{-\tau_{\rm ^{13}CO}}}
\int{T_{\rm mb}({\rm ^{13}CO}){\rm dv}}
\label{integ13}
\end{equation}
\begin{equation}
  \tau_{\rm C^{18}O} = - ln\left(1 - \frac{T_{mb}({\rm C^{18}O})}{15.81\left[\frac{1}{e^{15.81/T_{ex}}-1} - 0.0028\right]}\right)    
\label{tau18}
\end{equation}
\begin{equation}
{\rm N(C^{18}O)} = 8.26 \times 10^{13}~e^{\frac{15.80}{T_{ex}}}\frac{T_{ex} + 0.88}{1 - e^{\frac{-15.81}{T_{ex}}}} 
\int{\tau_{\rm C^{18}O}{\rm dv}}
\label{N18}
\end{equation}
with 
\begin{equation}
\int{\tau_{\rm C^{18}O}{\rm dv}} = \frac{1}{J(T_{ex}) - 0.045} \frac{\tau_{\rm C^{18}O}}{1-e^{-\tau_{\rm C^{18}O}}}
\int{T_{\rm mb}({\rm C^{18}O}){\rm dv}}
\label{integ18}
\end{equation}
The $J(T_{ex})$ parameter is $\frac{15.87}{exp(\frac{15.87}{T_{ex}}) - 1}$ in the case of Eq.\,(\ref{integ13}) and 
$\frac{15.81}{exp(\frac{15.81}{T_{ex}}) - 1}$ in Eq.\,(\ref{integ18}).
In all equations $T_{\rm mb}$ is the peak main brightness temperature obtained from Gaussian fits (see Table\,1-at the end of the manuscript-) 
and $T_{ex}$ the excitation temperature. Assuming that the \2 J=3--2 emission is optical thick the $T_{ex}$ was derived
from:
\begin{equation}
T_{ex} = \frac{16.6}{{\rm ln}[1 + 16.6 / (T_{\rm peak}(^{12}{\rm CO}) + 0.036)]}
\label{tex}
\end{equation}
where $T_{\rm peak}(^{12}{\rm CO})$ is the peak main brightness temperature obtained from the Gaussian fitting to 
the \2 J=3--2 line.
In cases that the \2 emission appears self-absorbed, the central component of the spectrum was corrected for absorption 
(see Fig.\,\ref{absCorrect}) in order to obtain a value for $T_{\rm peak}(^{12}{\rm CO})$. In those cases, it was used
the best single Gaussian that fits the wings of the self-absorbed
profile. This procedure was also applied in 
a few \3 spectra that presented signatures of self-absorption. It is important to remark
that the \3 and \8 spectra were carefully inspected to look for signatures of saturation in the line which would generate an
underestimation in some of its parameters. Besides the mentioned few cases of self-absorbed \3 emission, we did not find any feature 
suggesting line saturation in the \3 and \8 spectra. Line saturation is discussed in Sect.\,\ref{sat}.

Table\,2-at the end of the manuscript- presents the results obtained for each source: the source number, type, galactocentric distance,
and the integrated line ratio (I$^{13/18} = \int{T_{mb}^{13}~\rm{dv}}/\int{T_{mb}^{18}~\rm{dv}}$) 
are presented in Cols.\,1, 2, 3, and 4, respectively, and the T$_{ex}$, $\tau^{13}$, N(\3), $\tau^{18}$, N(\8), and the abundance ratio 
X$^{13/18}$ obtained from N(\3)/N(\8) are included in the others columns, respectively.

\begin{figure}[h]
\centering
\includegraphics[width=6.5cm]{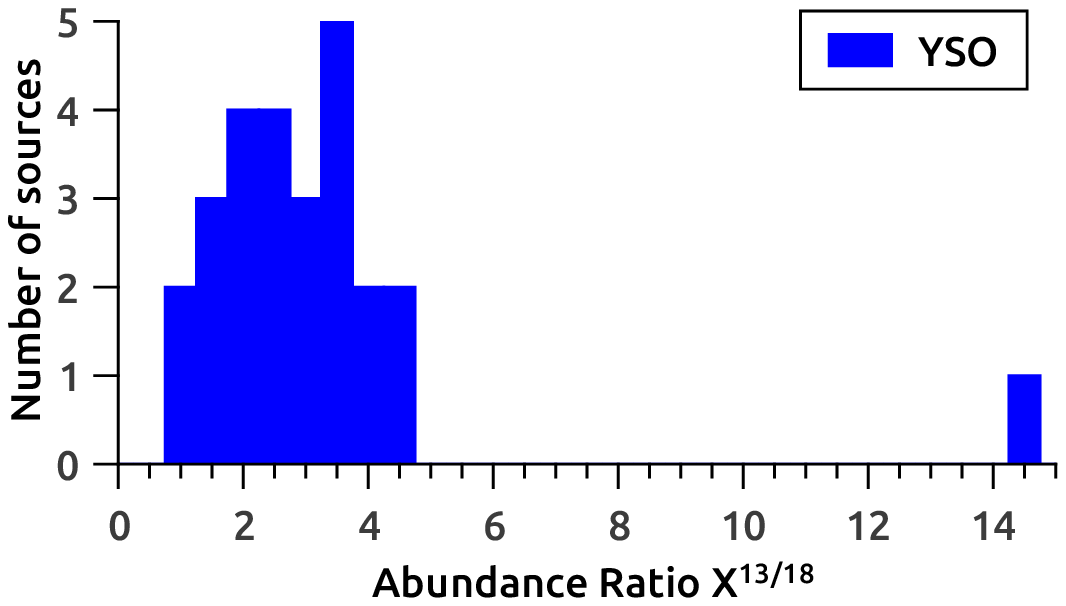}
\includegraphics[width=6.5cm]{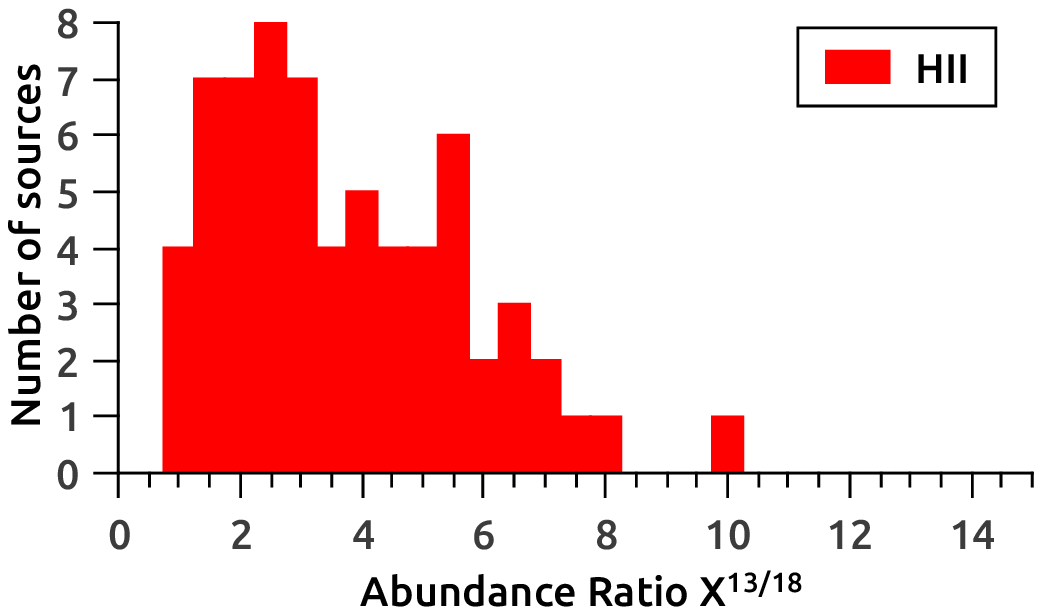}
\includegraphics[width=6.5cm]{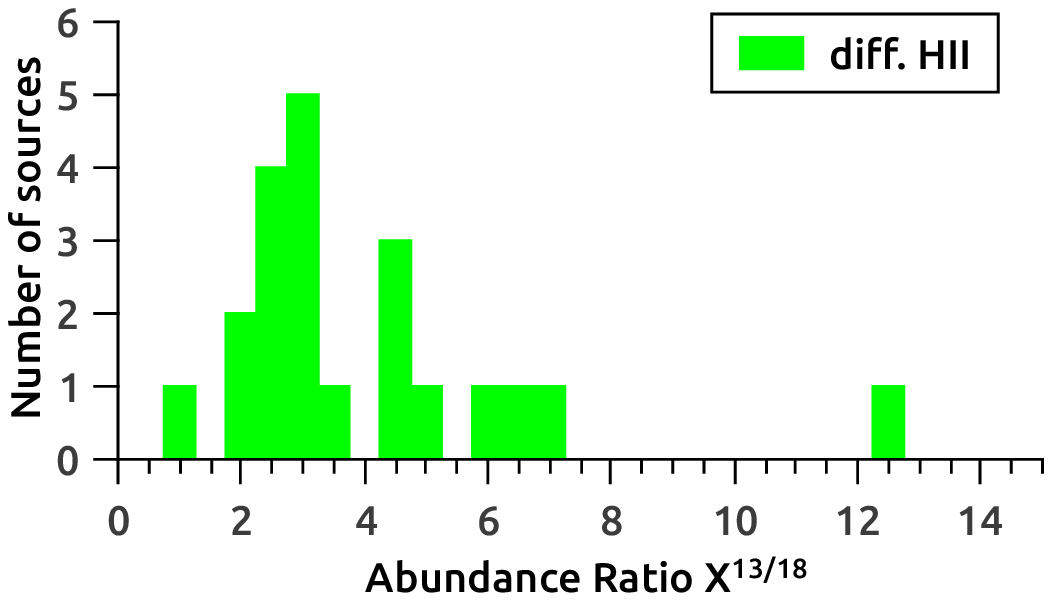}
\caption{Amount of sources vs. the \3/\8 abundance ratio (X$^{13/18}$). }
\label{histo}
\end{figure}

\begin{figure}[h]
\centering
\includegraphics[width=9.3cm]{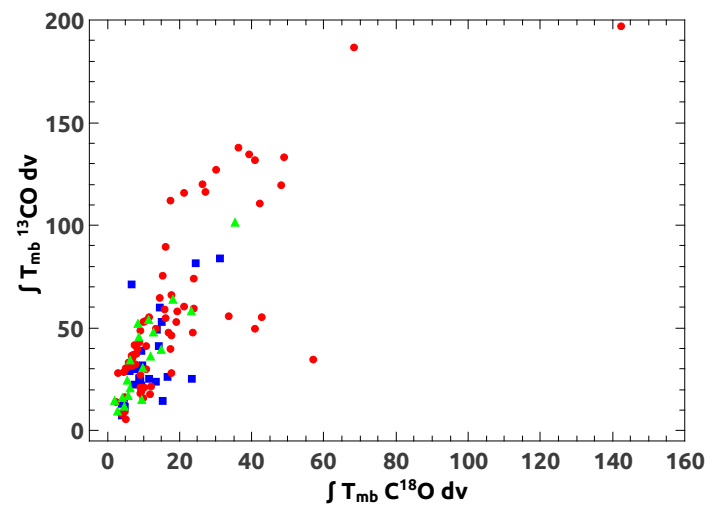}
\caption{Integrated \3 line vs. the integrated \8 line. YSOs, \hii~regions, and
diffuse \hii~regions are represented with blue squares, red circles, and green triangles, respectively.
}
\label{II}
\end{figure}

Figure\,\ref{histo} displays in histogram diagrams the number of sources vs. the abundance ratio (X$^{13/18}$) for
YSOs, \hii~regions, and diffuse \hii~regions, respectively. The width of the histograms bar is
0.5, thus, the height of the bar represents the amount of sources that have abundance ratios
distributed from the center of the bar $\pm 0.25$.
Figure\,\ref{II} shows the integrated line relation between the isotopes, i.e. I$^{13} = \int{T_{mb}^{13}~\rm{dv}}$ vs.
I$^{18} = \int{T_{mb}^{18}~\rm{dv}}$.

Table\,\ref{tableAve} presents the mean values of the abundance and integrated line ratios ($\overline{\rm X^{13/18}}$ and
$\overline{\rm I^{13/18}}$) for each kind of source together with the total number of sources composing each sample (N).
Values between brackets are the results obtained after removing the outliers points of each sample (see the bars totally 
detached from the main distribution in Fig.\,\ref{histo}). 

\begin{figure}[h]
\centering
\includegraphics[width=9.3cm]{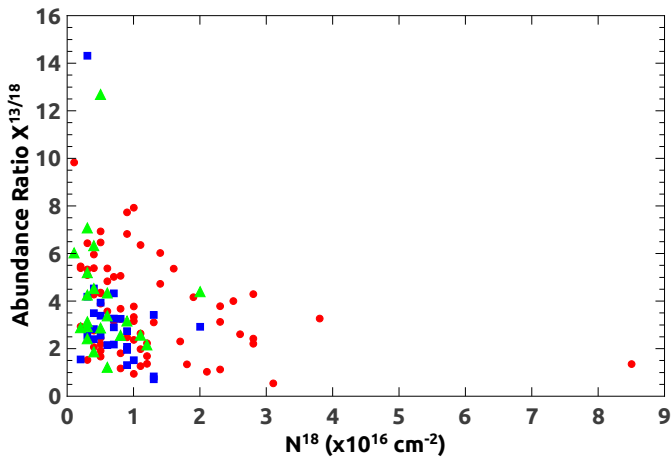}
\caption{Abundance ratio vs. \8 column density. YSOs, \hii~regions, and
diffuse \hii~regions are represented with blue squares, red circles, and green triangles, respectively. }
\label{XN}
\end{figure}

\begin{figure}[h]
\centering
\includegraphics[width=9.3cm]{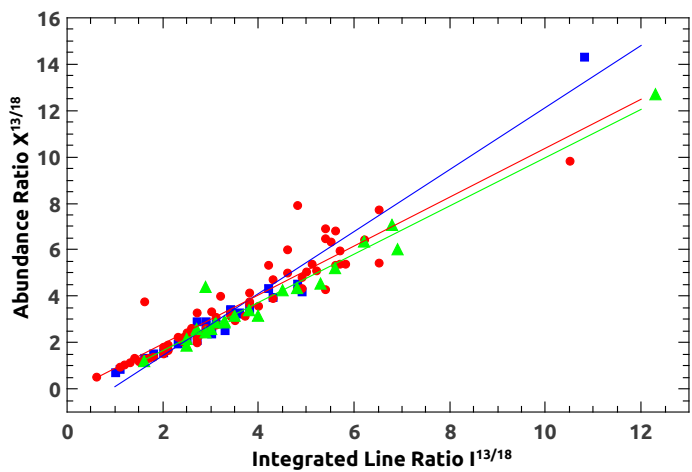}
\caption{Abundance ratio X$^{13/18}$ vs. integrated line ratio I$^{13/18}$. YSOs, \hii~regions, and
diffuse \hii~regions are represented with blue squares, red circles, and green triangles, respectively. Linear fittings
to each set of data for the whole sample are displayed. }
\label{XI}
\end{figure}

Figure\,\ref{XN} displays the obtained abundance ratio for each source vs. the \8 column density.
Figure\,\ref{XI} displays the abundance ratio X$^{13/18}$ vs. the integrated line
ratio I$^{13/18}$, which shows good linear relations. The results from linear fittings for each kind of source are presented in 
Table\,\ref{linefit}. It is also included the $\chi^{2}$ factor obtained from each fitting. As in Table\,\ref{tableAve}, between
brackets it is presented the results obtained after removing the above mentioned outliers points. These points are: 
(X$^{13/18}$, I$^{13/18}$) $=$ (14.3, 10.8)
for a YSO, (9.8, 10.5) for an \hii~region, and (12.7, 12.3) for a diffuse \hii~region.

\begin{table}
\caption{Mean values of the abundance and integrated line ratios.}
\label{tableAve}
\centering
\begin{tabular}{lccc}
\hline
 &               N & $\overline{\rm X^{13/18}}$ & $\overline{\rm I^{13/18}}$  \\
\hline
YSO &              26  &  3.13 (2.68) & 3.27 (2.97)   \\
\hii &             67  &  3.68 (3.59) & 3.65 (3.55)  \\
diffuse \hii  &    21  &  4.09 (3.66) & 4.36 (3.96)  \\
\hline
\multicolumn{4}{l}{\tiny {\it Note:} Values between brackets are the results obtained}\\ 
\multicolumn{4}{l}{\tiny after removing the outliers points of each sample. }\\ 
\end{tabular}
\end{table}

\begin{table}
\caption{Linear fitting results $(A~x + B)$ from the data presented in Fig.\,\ref{XI}.}
\tiny
\label{linefit}
\centering
\begin{tabular}{lccc}
\hline
               & $A$             &  $B$                  & $\chi^{2}$  \\
\hline
YSO            & 1.33$\pm$0.05 (0.98$\pm$0.04)  &  $-$1.24$\pm$0.19 ($-$0.25$\pm$0.15)    &   0.96 (0.94)      \\
\hii           & 1.05$\pm$0.05 (1.10$\pm$0.05)  &  $-$0.17$\pm$0.20 ($-$0.32$\pm$0.22)    &   0.87 (0.84)     \\
Diffuse \hii   & 1.04$\pm$0.04 (0.97$\pm$0.07)  &  $-$0.43$\pm$0.24 ($-$0.20$\pm$0.30)    &   0.95 (0.89)     \\
\hline
\multicolumn{4}{l}{\tiny {\it Note:} Values between brackets are the results obtained after removing the outliers }\\          
\multicolumn{4}{l}{\tiny points of each sample. }\\      
\end{tabular}
\end{table}

\section{Discussion}

Taking into account that the molecular gas related to YSOs should be less affected by the UV radiation than the gas associated 
with \hii~regions, from the analysis of the \3/\8 abundance ratios X$^{13/18}$ obtained towards the 114 studied sources, as a 
first result, it can be suggested that the X$^{13/18}$ increases as the degree of UV radiation increases. 
It is important to note that it is likely that the molecular gas related to some sources catalogued as YSOs can be affected by the 
UV radiation. These sources could be located at the borders of extended \hii~regions, and/or they may be transiting the last stages 
of star formation and have begun to ionize their surroundings. However, we consider that the gas associated with most of them should be
less affected by the radiation than in the case of sources catalogued as \hii~and diffuse \hii~regions. 
This phenomenon is indeed reflected 
in our analysis. Figure\,\ref{histo} shows that YSOs tend to have smaller X$^{13/18}$ values than the other type of sources, 
which can be also appreciated by comparing the mean values presented in Table\,\ref{tableAve}.
This result is in agreement with what it is observed in different regions of molecular clouds that are affected by far-UV radiation, 
such as the Orion-A giant molecular cloud \citep{shima14}, and what it is predicted from photodissociation models \citep{visser09,van88}.
Thus, we confirm, through a large sample of sources, that selective far-UV photodissociation of \8 indeed occurs. 
Moreover, as Fig.\,\ref{XN} shows, the X$^{13/18}$ ratio decreases with the increasing \8 column density in all sources,
which suggests that this phenomenon occurs even considering each group of source separately. 

The relation between the isotopes integrated line (I$^{13}$ vs. I$^{18}$, see Fig.\,\ref{II}) shows slight differences  
between the kind of source. This is also reflected in the relation between the abundance and the integrated line ratios (Fig.\,\ref{XI}), 
which is an interesting relation because it compares values that were derived from excitation considerations (LTE assumption) with 
values that are direct measurements. These relations show that the \3/\8 abundance ratio could be estimated directly from the integrated
line ratios, as it can be appreciated by comparing the values presented in Table\,\ref{tableAve}.

%This relation is linear and it changes slightly from one type of source to other. 
%It is observed a decay in the slope of the linear fittings with the increase in the evolutive stage of the source.
% Even though it would be useful to increase the sample of sources and hence the statistical analysis, 
%we propose that these relations could be used to convert measured isotopes integrated line ratios into abundance ratios
%in future works.

\subsection{Saturation of the J=3--2 line}
\label{sat}

It is known that a linear molecule may increase its opacity with the increasement of the 
J rotational level until reaching a J maximum (J$_{max}$), for which the optical depth exhibits a peak \citep{gold99}.
The J$_{max}$ depends on the temperature and the molecule rotational constant. Thus, depending on the temperature 
of the region, it is possible that the CO J=3--2 line can suffer more saturation than the lower transitions, and hence
a quantitative comparison of the abundance ratio with previous results obtained from the J=2--1 and 1--0 lines should be done 
with caution.
 
\citet{wout08} measured the \3/\8 integrated intensity ratio (I$^{13/18}$) using the J=1--0 and J=2--1 line towards some Galactic star forming 
and \hii~regions, and suggested that saturation can be more pronounced in the J=2--1 transition. 
We compare our \3/\8 integrated intensity ratios with those
obtained by \citet{wout08} (see their Fig.\,3) in the galactocentric distance range (4--8 kpc). We find that they do not present 
large discrepancies. Our results are very similar to those obtained from the J=2--1 line in \citet{wout08}, suggesting that the saturation
does not increase considerably from J=2--1 to J=3--2 in this kind of sources, and moreover, in some cases it is comparable even 
with the obtained from the J=1--0 line.

Thus, taking into account that the \3 and \8 spectra do not present any signature of line saturation, almost all $\tau^{13}$, 
and all $\tau^{18}$ values are lower than unity (see Table\,2-at the end of the manuscript), and our 
\3/\8 integrated intensity ratios are in good agreement with those obtained from the J=2--1 and 1--0 lines in a previous work towards 
similar sources, we conclude that line saturation should not be an important issue in our analysis.

\subsection{Abundance ratio and the distance}

\begin{figure}[h]
\centering
\includegraphics[width=9.2cm]{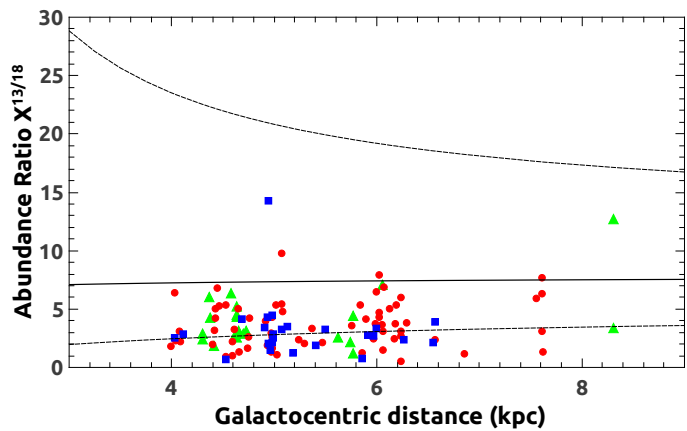}
\caption{Abundance ratio X$^{13/18}$ vs. galactocentric distance. The 
continuous black curve is the plot of the \3/\8 abundance ratio from \citet{wilson94} (Eq.\,\ref{ratioEq}). 
Dashed curves are the result of plotting this equation considering the errors bars.  }
\label{Xdist}
\end{figure}

Given that the elemental abundance ratios $^{12}$C/$^{13}$C, $^{16}$O/$^{18}$O, between others, 
presented in \citet{wilson94} are largely used in the literature when CO data are studied,
we compare our results with them.
The abundance relations presented in \citet{wilson94} are:
\begin{gather}
{\rm (^{12}C/^{13}C) = (7.5 \pm 1.9) \times D_{GC} + (7.6 \pm 12.9)}\\
{\rm (^{16}O/^{18}O) = (58.8 \pm 11.8) \times D_{GC} + (37.1 \pm 82.6)}
\label{abundWil}
\end{gather}
where D$_{\rm GC}$ is the galactocentric distance. Thus, to assume a 
\3/\8 abundance ratio one must to perform a double ratio between the above expressions, yielding:
\begin{equation}
{\rm \frac{^{13}CO}{C^{18}O} = \frac{(58.8 \pm 11.8) \times D_{GC} + (37.1 \pm 82.6)}{(7.5 \pm 1.9) \times D_{GC} + (7.6 \pm 12.9)}}$$
\label{ratioEq}
\end{equation}
Figure\,\ref{Xdist} displays the abundance ratio X$^{13/18}$ vs. galactocentric distance for the sources analyzed
in this work. The black curve is the plot of Eq.\,\ref{ratioEq}, which shows that the mean value of the X$^{13/18}$ ratio
is between 7 and 8 for the whole galactocentric distance range. The dashed curves are the result of plotting
this equation considering the errors bars. These curves delimit a region 
where, according to \citet{wilson94}, the sources should lie.

The analyzed sources lie mainly in the galactocentric distance range 4.0--6.5 kpc, and a few has farther
distances, around 8 kpc. Thus our analysis lack of information concerning to the molecular gas lying
within a radius of 4 kpc from the Galactic center. From Fig.\,\ref{Xdist} it can be appreciated that almost all the abundance ratio
values of our complete sample are lower than the mean value predicted by \citet{wilson94}, and moreover, there are many values lying under 
the lower limit from Eq.\,\ref{ratioEq} (bottom dashed curve in Fig.\,\ref{Xdist}). This suggests that the \3/\8 abundance ratio 
derived from the double ratio between $^{12}$C/$^{13}$C and $^{16}$O/$^{18}$O could overestimate the actual value.

\section{Summary and concluding remarks}

Using the \2, \3, and \8 J=3--2 emission obtained from the COHRS and CHIMPS surveys performed with the JCMT telescope 
and using additional data from the telescope database, we studied the \3/\8 abundance ratio towards a large sample of YSOs 
and \hii~regions located in the first Galactic quadrant. 

From the statistical analysis of the X$^{13/18}$ ratio we found that YSOs have, in average, smaller values 
than \hii~and diffuse \hii~regions. Taking into account that the gas associated with YSOs should be
less affected by the radiation than in the case of \hii~and diffuse \hii~regions, we can confirm the selective far-UV 
photodissociation of \8 as it was observed in previous works towards particular molecular clouds and as it was predicted by models.
Additionally, it was found a linear relation between the abundance ratios and the integrated line ratios (I$^{13/18}$),
suggesting that the \3/\8 abundance ratio could be estimated directly from I$^{13/18}$. 
%In future studies of similar sources that make use of \3 and \8 observations and lacks of a T$_{ex}$ value to determine column 
%densities it could be used this relation to estimate an abundance ratio.

Most of the sources are located in the galactocentric distance range 4.0--6.5 kpc, which indicates that we are mainly studying 
sources in the Scutum-Crux and Sagitarius-Carina Galactic arms. A few sources are located so far as 8 kpc. 
Thus our analysis lack of information concerning to the molecular gas lying within a radius of 4 kpc from the Galactic center.
Extension of the used surveys, or others \2, \3, and \8 surveys covering the inner Galaxy would be useful to complete this study.
From the covered galactocentric distance range it was shown that the \3/\8 abundance ratios obtained directly from
the molecular emission are lower than if they are derived from the known elemental abundances relations.

Finally, it is important to mention that this is the first \3/\8 abundance ratio study obtained directly from CO observations
towards a large sample of sources of different nature at different locations. 
Thus, the \3/\8 abundance ratios derived in this work would be useful for future studies of molecular gas associated with YSOs and \hii~regions 
based on the observation of these isotopes.

\begin{figure*}[h!]
\centering
\includegraphics[width=9cm,angle=-90]{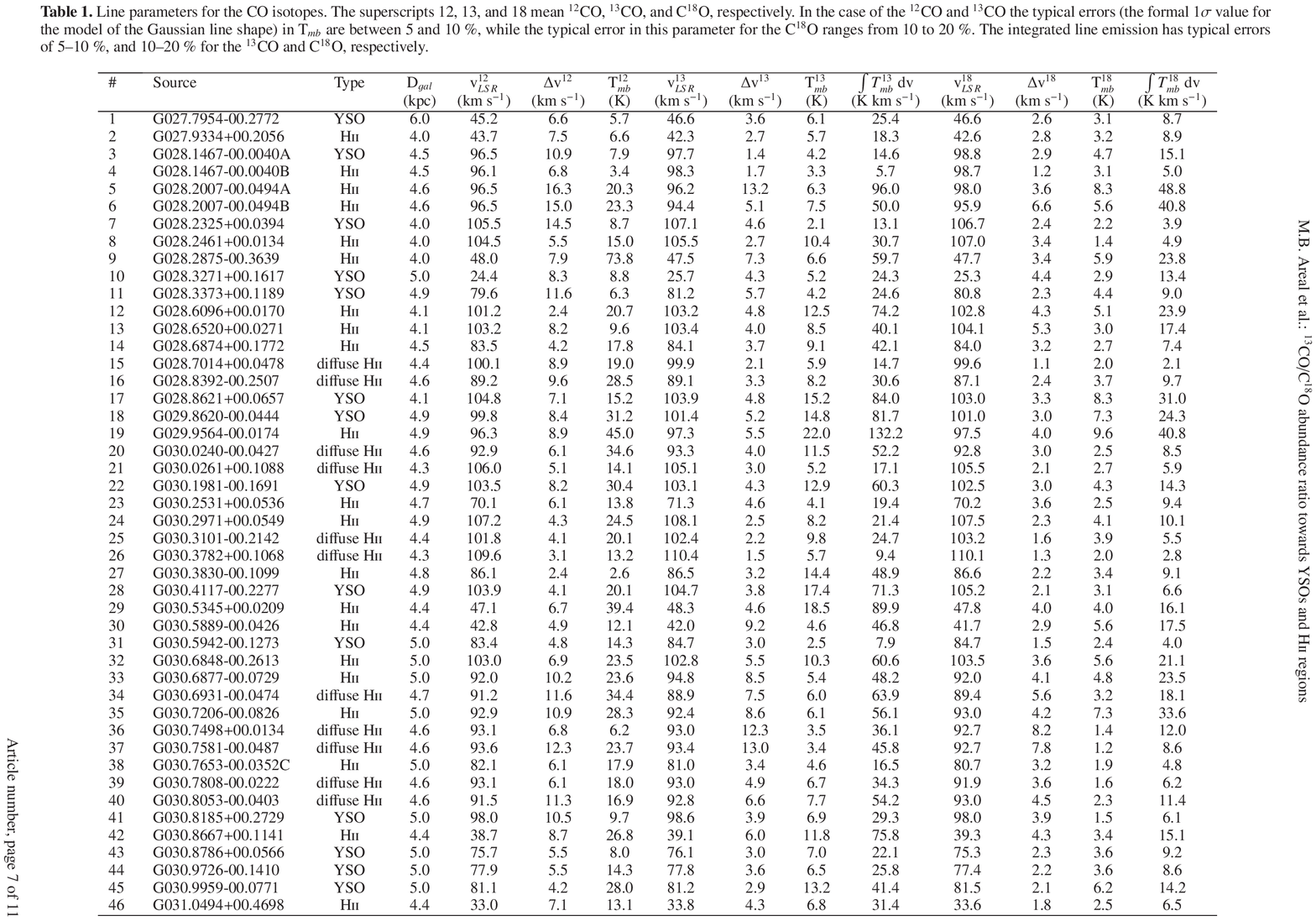}
\includegraphics[width=9cm,angle=-90]{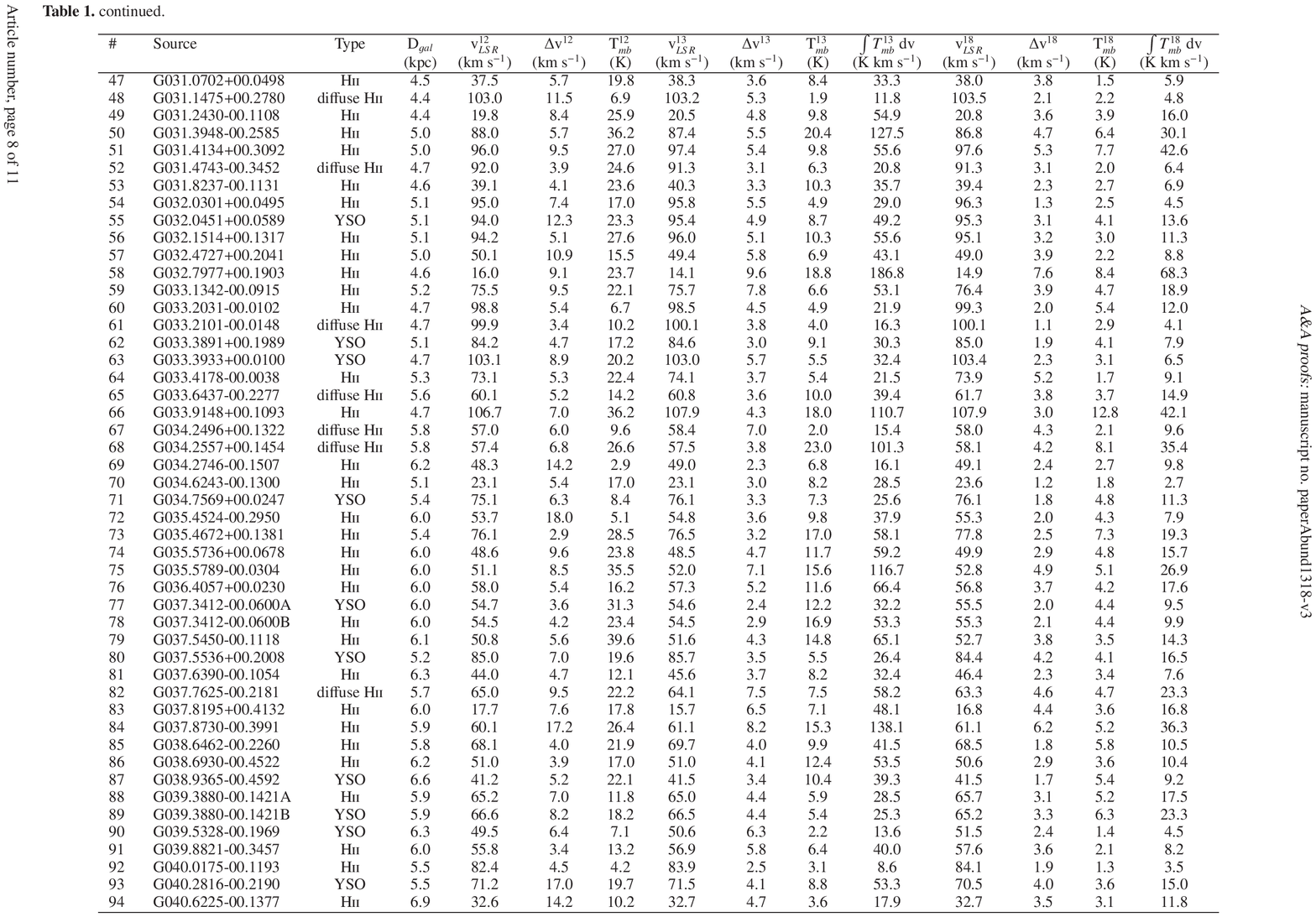}
\includegraphics[width=9cm,angle=-90]{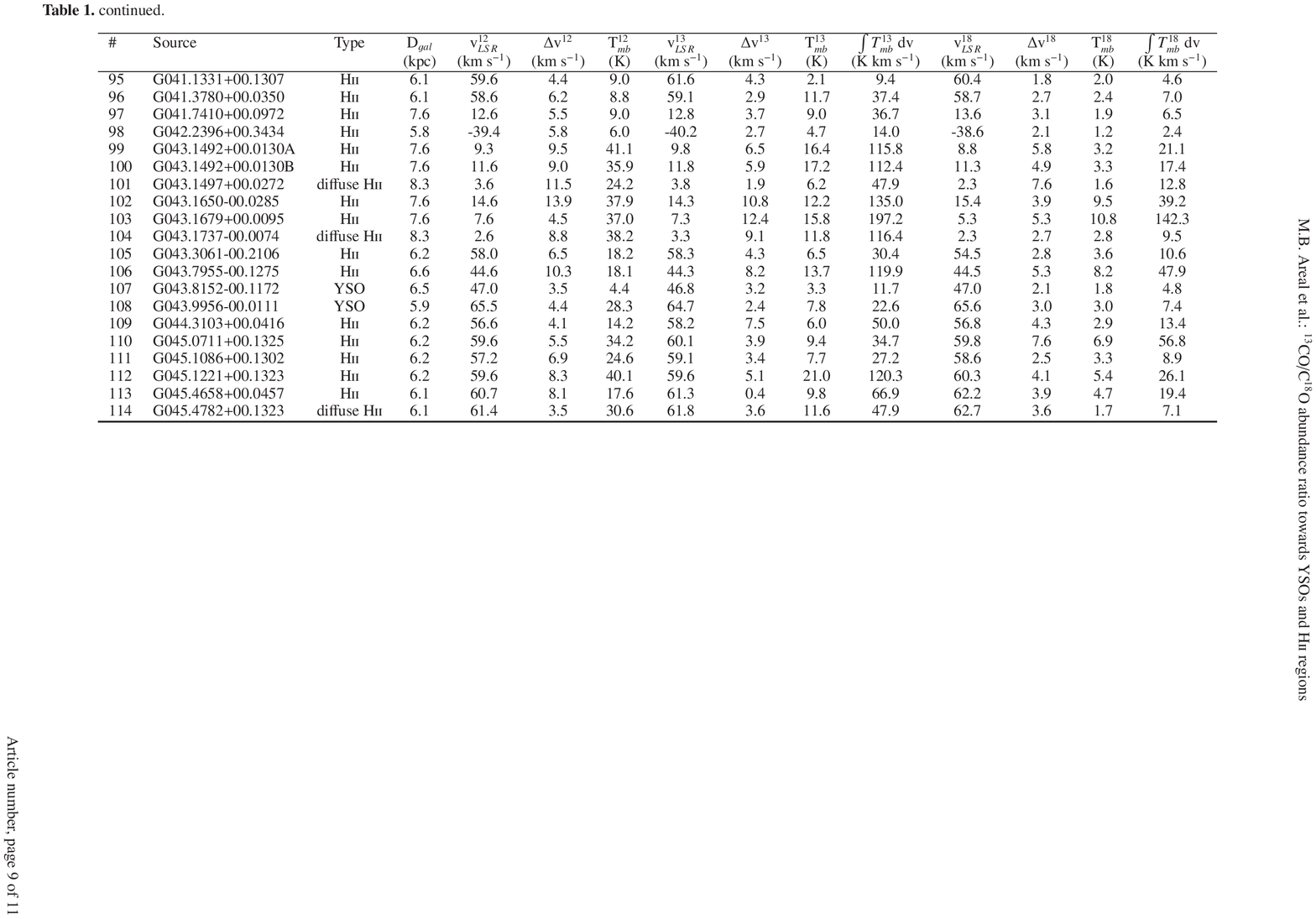}
\caption{Table1}
\end{figure*}

\begin{figure*}[h!]
\centering
\includegraphics[width=9cm]{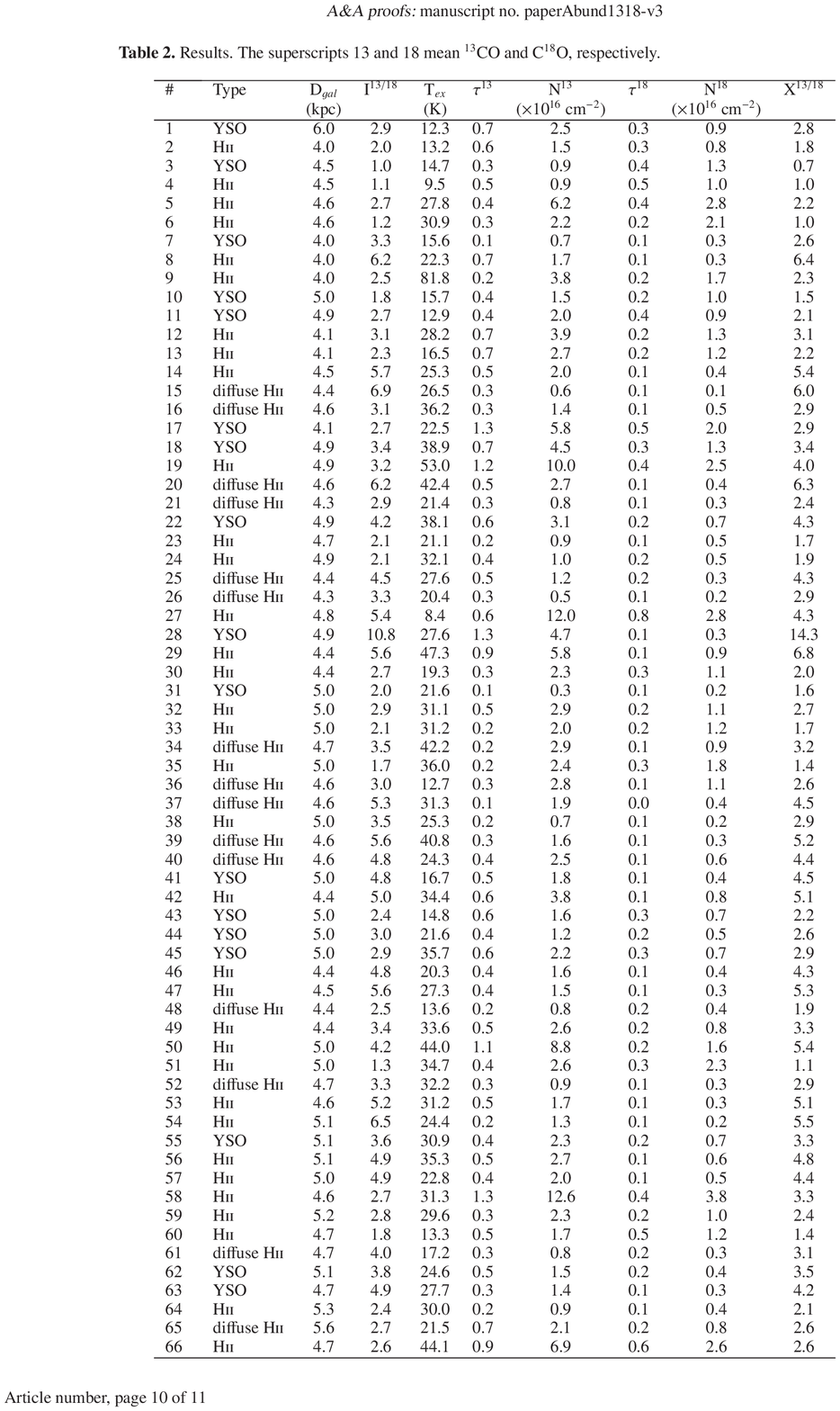}
\includegraphics[width=9cm]{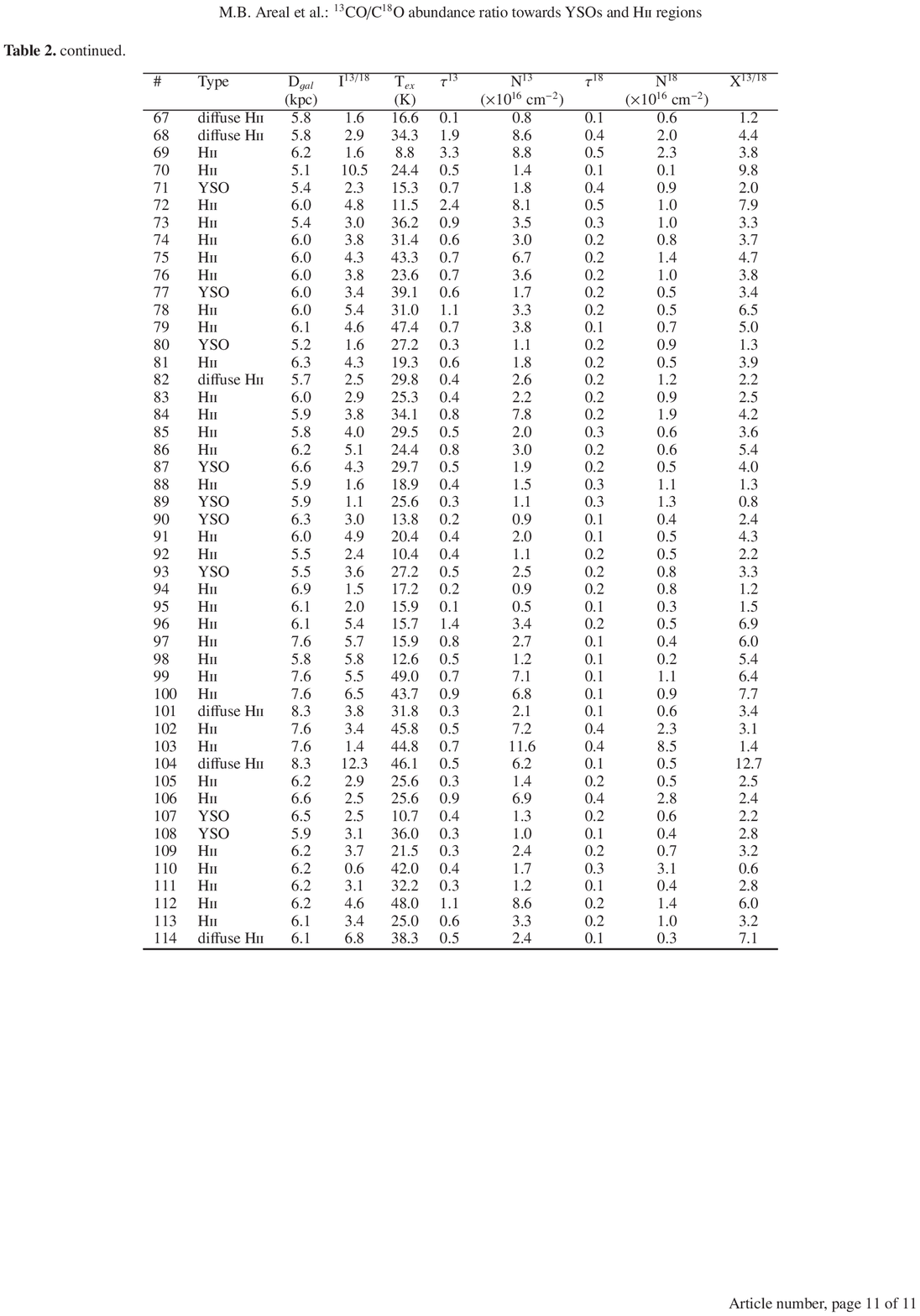}
\caption{Table2}
\end{figure*}

\section*{Acknowledgments}

We thank the anonymous referee for her/his helpful comments and suggestions.
M.B.A. and M.C.P. are doctoral fellows of CONICET, Argentina. 
S.P. and  M.O. are members of the {\sl Carrera del Investigador Cient\'\i fico} of CONICET, Argentina. 
This work was partially supported by Argentina grants awarded by UBA (UBACyT), CONICET and ANPCYT.

%%%%%%%%%%%%%%%%%%%%%%%%%%%%%%%%%%%%%%%%%%%%%%%%%%%%%%%%%%%%%%%%%%%%%
\bibliographystyle{aa}  % A&A format
   %\bibliographystyle{klunamed}     
   % format of references provided by the review (.bst)
\bibliography{ref}
   % file containing the bibtex references (.bib)
\IfFileExists{\jobname.bbl}{}
{\typeout{}
\typeout{****************************************************}
\typeout{****************************************************}
\typeout{** Please run "bibtex \jobname" to optain}
\typeout{** the bibliography and then re-run LaTeX}
\typeout{** twice to fix the references!}
\typeout{****************************************************}
\typeout{****************************************************}
\typeout{}
}

\label{lastpage}
\end{document}